\documentclass[11pt,a4paper]{article}
\pdfoutput=1
\usepackage{jcappub}
\usepackage{sirimacro}

\usepackage{amsmath}
\usepackage{amsfonts}
\usepackage{amssymb}

\usepackage{graphicx}
\usepackage{epsfig}

\usepackage{mathrsfs}
\usepackage{bm}
\usepackage{microtype}

\usepackage{hyperref}

\begin{document}

\title{On the Abundance of Extreme Voids}

\author{Sirichai Chongchitnan}

\affiliation{E. A. Milne Centre for Astrophysics, University of Hull,  Cottingham Rd., Hull, HU6 7RX, United Kingdom.}

\emailAdd{s.chongchitnan@hull.ac.uk}

\abstract{Cosmic voids have been shown to be an effective probe of cosmology, complementary to galaxy clusters. In this work, we present a simple theoretical framework for  predicting of the size of the largest voids expected within a given redshift and volume. Our model is based on the exact extreme-value statistics which has previously been successfully applied to massive galaxy clusters. We implement our formalism using the void-abundance models and compare the extreme-void predictions to simulations and observations. We find that the simplest void models can only explain the extreme-void abundance with ad hoc parameter adjustments. We argue that extreme-void distributions should be used as an additional test on theories of void abundance.}

\maketitle

\section{Introduction}
In the standard concordance ($\Lambda$CDM) cosmology, cosmic structures on the largest scales consist primarily of massive galaxy clusters (typically of masses $\sim10^{15}M_\sun$) and large cosmic voids of radius $\sim100$ Mpc. Thanks to ambitious ground and space-based surveys and a suite of large $N$-body simulations, the distribution of massive galaxy clusters have now been studied in great detail. Several analytic and semi-analytic mass functions are now available to describe and predict cluster distribution \cite{ps,sheth,tinker,warren,reed}. These prescriptions have generally been shown to be in good agreement with simulations and observations.

Cosmic voids, on the other hand, have not received the same level of attention as clusters. The main difficulty is the lack of consensus of what defines a void, for instance, the underdensity threshold below which a region is said to be a void. Furthermore, whilst clusters can be observed via the mass-luminosity relation, true cosmic voids are empty of dark matter, and so cannot be directly observed. The relation between galaxy voids and dark matter voids is still not completely understood (see \cite{leclercq,sutter} for recent progress). 

Nevertheless, a number of recent simulations have shown convincingly that voids, too, are a sensitive probe of cosmology, because they preserve much of the conditions in the early Universe. Using indirect probes such as gravitational lensing and the Integrated Sachs-Wolfe effect, their abundance, shape and density profile have been shown to be effective probes of dark energy \cite{park,bos,chantavat}, dark matter \cite{hellwing,yang}, gravity theories \cite{cai,clampitt,zivick} and primordial non-Gaussianity \cite{me,lavaux}.

The first convincing analytic model of void abundance was given by Sheth and Van de Weygaert (SVdW) \cite{svdw}. Although it has since undergone minor modifications by a number of previous authors \cite{jennings, furlanetto, paranjape, achitouv}, the heart of the theory, based on excursion-set formalism, remains largely unaltered.  In this work, we ask: how large are the largest voids predicted by the SVdW theory? Our focus on the largest voids is based on the fact that they have the most interesting phenomenology, and may be responsible for large-scale anomalies in the CMB anisotropies, particularly the notorious Cold Spot \cite{inoue, finelli, szapudi, nadathur}, and may produce similar anomaly in the 21cm fluctuations \cite{kovetz}.

At the core of our work is the exact extreme-value statistics previously used for predicting the distribution of the most massive clusters \cite{harrison1,harrison2}. We will use this framework, together with the SVdW theory, to produce a probability distribution of the largest voids expected to be observed at a given redshift. Our framework is then tested against simulation and observational data. We then discuss how this approach can be improved and extended to other void models.

Even though simulations give us access to the full distribution of void number counts, the results are highly dependent on the choice of void finder (\eg{} ZOBOV \cite{zobov}, VIDE \cite{vide}, amongst others, see \cite{nada2} for a recent discussion). Our goal is in this work is to investigate whether the extreme-void predictions from current theory is consistent with the extreme voids seen in simulations and observations. And if not, how could we address the inconsistency?
 
Throughout this paper, we will use the Planck+WMAP cosmological parameters as given in \cite{lahav}.

\section{Void abundance}
 

We begin with an overview of the SVdW theory, which gives the differential abundance of voids in the radius interval $[R,R+dR]$. The model is based on the so-called `excursion set' theory, in which the local linear overdensity, $\delta(\mb{x})$, performs a random walk in the $(\delta, \sigma^2)$ plane, where $\sigma^2$ is the variance of the overdensity, smoothed on comoving scale $R$:
\ba\sigma^2(R,z) = D(z) \int_0^\infty {\D k\over 2\pi^2} k^2 P(k) W^2(k,R).\ea
$D(z)$ is the amplitude of the growing mode (normalised to $D(0)=1$), $P(k)$ is the linear matter power spectrum and $W(k,R)$ is the Fourier-space window function.  A void of radius $R$ is deemed to form when $\delta$ first makes an `excursion' below the barrier $\delta_v$. In the original paper, the value $\delta_v=-2.7$ was recommended (this corresponds to a nonlinear overdensity $\Delta_v \approx-0.8$). This value of $\delta_v$ is derived from the `shell-crossing' condition, analogous to the Einstein-de Sitter critical overdensity, $\delta_c=1.686$, commonly used to define an object formed by spherical collapse.


The SVdW model gives the differential abundance of voids with radius $R$ as \cite{svdw}
\ba
\diff{n}{\ln R}&= {f(\sigma)\over V_s(R_L)}  \diff{\ln\sigma^{-1}}{\ln R_L}\bigg|_{R_L = R/1.7}, \\
f(\sigma)&=2\pi x^2\sum_{j=1}^\infty  j e^{-(j\pi x)^2/2}\sin(j\pi\mc{D}),\\
\mc{D}&= \bkt{1+ \delta_c/|\delta_v|}^{-1}, \qquad x=\mc{D}\sigma/|\delta_v|. \notag\ea
where $V_s(r)=4\pi r^3/3$.  The oscillatory behaviour of $f$ at high $R$ can be avoided with the approximation given by Jennings, Li and Hu \cite{jennings}:
\be f(\sigma)\approx\begin{cases} \sqrt{2\over\pi}{|\delta_v|\over\sigma}e^{-\delta_v^2/2\sigma^2},\quad &x\leq0.0276,\\ 2\pi x^2\sum_{j=1}^\infty  j e^{-(j\pi x)^2/2}\sin(j\pi\mc{D})\quad &\mbox{otherwise}.\end{cases}\ee

The SVdW theory has been shown to be consistent with dark-matter simulations where only spherical voids are considered \cite{jennings}. However, void finders based on the watershed algorithm typically pick out voids that are non-spherical and irregular. This led many authors to take $\delta_v$ as a free parameter in the model, and simply adjust $\delta_v$ to match simulation results. Whilst this is a rather \ii{ad hoc} approach, such an adjustment appears to do the trick, for instance, in accurately matching the void abundances in the simulations of \cite{chan}, with the parameter $\delta_v\approx-1$ for $z\lesssim1$. We will return to these simulation results in \S\ref{sim}.


\section{Void number counts}
The probability density function (pdf) for voids with radius in the interval $[R,R+dR]$, in the redshift bin centred $z$, width $\Delta z$, is given by
\ba
f(R,z)= {f\sub{sky}\over N\sub{tot}}  \int^{z+\Delta z/2}_{z-\Delta z/2} \textrm{d}z  \diff{V}{z} {1\over R}\diff{n}{\ln R},\lab{pd}
\ea
where $f\sub{sky}$ is the fraction of the sky observed and $dV/dz$ is the volume element given by
\ba\diff{V}{z}&=f\sub{sky} {4\pi\over H(z)}\bkt{\int_0^z{\D z^\pr\over H(z^\pr)}}^2, \\
H(z)&\approx H_0\bkts{\Omega_m(1+z)^3+\Omega_\Lambda}^{1/2}.
\ea 
In \re{pd}, the total number of voids, $N\sub{tot}$, with radius above $R\sub{min}$ in the redshift bin $[z-\Delta z/2, z+\Delta z/2]$ is
\ba
N\sub{tot}(z)= f\sub{sky} \int_{R\sub{min}}^{R\sub{max,V}}{\textrm{d}R \over R}\int^{z+\Delta z/2}_{z-\Delta z/2} \textrm{d}z  \diff{V}{z} \diff{n}{\ln R}.
\ea
where $R\sub{max,V}$ is determined by the volume of the redshift bin.

\begin{figure} 
   \centering
   \includegraphics[height=4.05in]{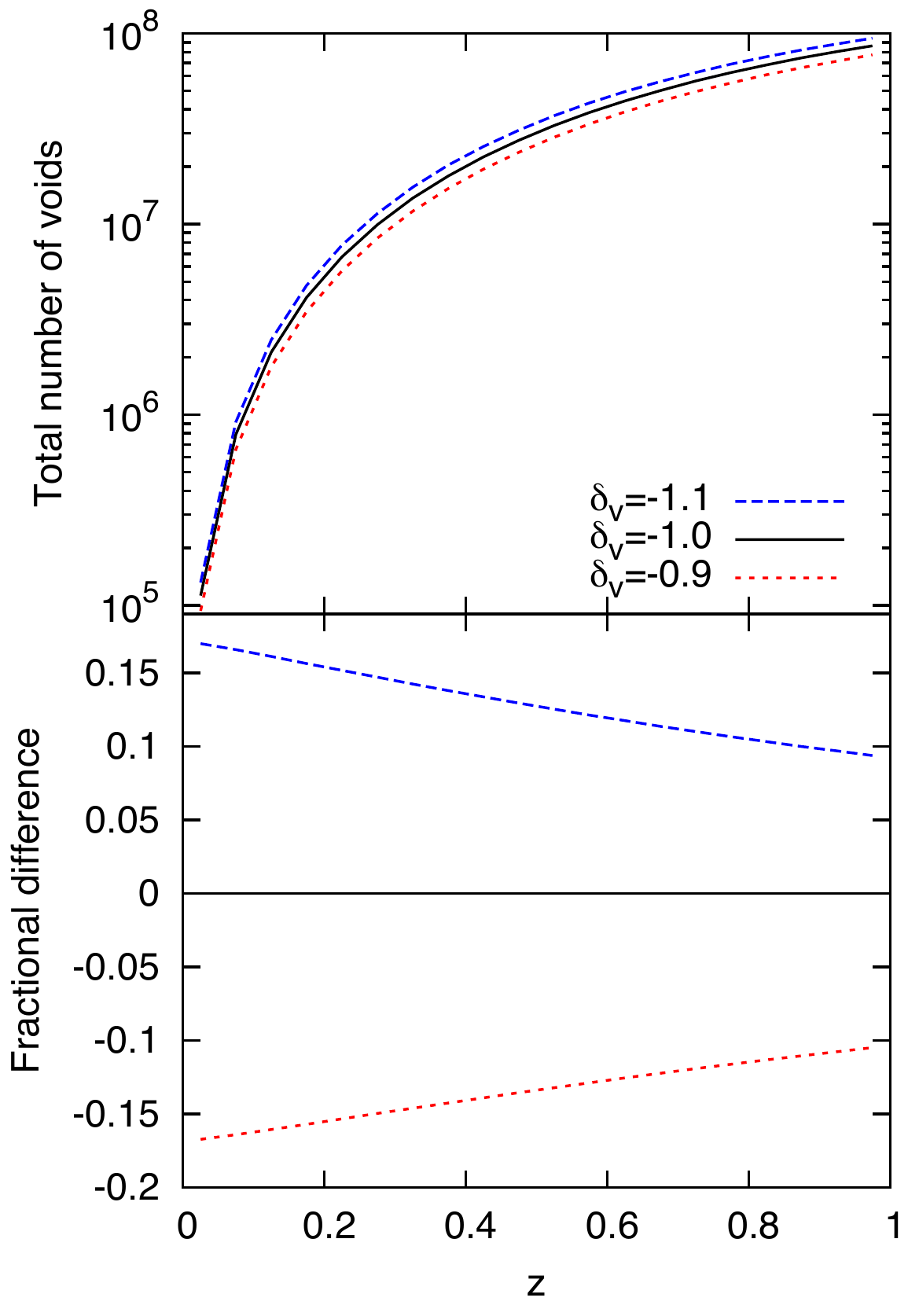}  
   \caption{The top panel shows total number of voids $N\sub{tot}$ and its variation with respect to the void underdensity $\delta_v$. The lower panel shows the fractional difference when compared with the fiducial model with $\delta_v=-1$.}
   \label{figNtot}
\end{figure}

Fig. \ref{figNtot} shows total number of voids $N\sub{tot}$ as a function of $z$ (top panel), with  $\Delta z=0.05$, $R\sub{min}=1$ and $f\sub{sky}=1$. Since we will be concerned with the largest voids, our results are insensitive to changes in $R\sub{min}$. In the same plot, we consider what happens when $\delta_v$ varies from $-0.9$ to $-1.1$, and the fractional difference compared to the $\delta_v=-1$ model is plotted in the lower panel. 

We see that when the void barrier is decreased (voids become emptier), the number count increases (in this case by about $10-15\%$), and vice versa. This can be interpreted as the fragmentation within the void population -- smaller voids are typically emptier, and they can combine to form shallower, larger voids, which are therefore less numerous. This simple picture is consistent with detailed studies of void profiles in previous simulations \cite{sutter,cai}.




Having obtained the pdf \re{pd}, we can construct the cumulative probability distribution (cdf), $F(R)$, which gives the probability that an observed void has radius $\leq R$. It is given by: 
\ba F(R)=\int_{R\sub{min}}^R f(r)\, \D r.\ea

\section{Distribution of extreme voids}\lab{distex}

Harrison and Coles \cite{harrison} showed that the cdf naturally leads to the prediction of the most massive clusters expected at a given redshift. This section follows their treatment, but adapted to cosmic voids.

Consider $N$  observations of voids drawn from the cdf, $F(R)$. We ask: what is the probability that the largest void observed is of radius $R^*$? The cdf required is
\ba \Phi(R^*,N)=F_1(R\leq R^*)\ldots F_N(R\leq R^*)=F^N(R^*),\ea
where the last step assumed that void radii are independent, identically distributed variables (we will come back to this point). Differentiating $\Phi$ with respect to $R^*$ gives the \ii{exact extreme-value} pdf
\ba \phi(R^*,N)=\diff{}{R^*}F^N(R^*)=Nf(R^*)[F(R^*)]^{N-1}.\lab{eev}\ea

\begin{figure} 
   \centering
   \includegraphics[height=1.9in]{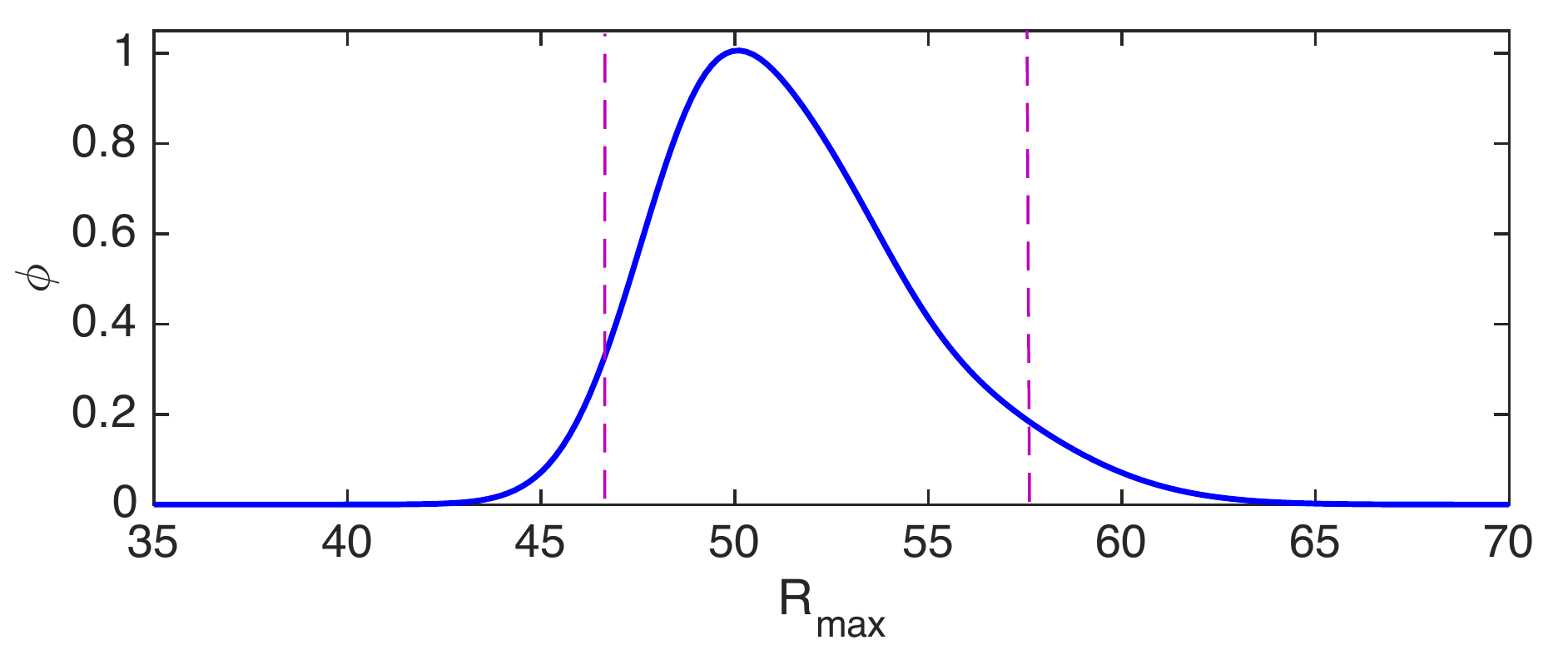}   
   \caption{The extreme-value pdf, $\phi$, at $z=0.225$. The values of the void radius, $R$, corresponding to the 5th (left) and 95th (right) percentiles are shown in dashed lines. }
   \label{fig_pdf}
\end{figure}

Figure \ref{fig_pdf} shows an example of the exact extreme-value pdf, $\phi(R)$, at $z=0.225$. The pdf is non-Gaussian but can be well-approximated as a sum of a few Gaussian terms. On the figure we also show the contour in dashed lines representing the 5th and 95th percentile of the distribution.

\no Assume that all available voids are observed ($N=N\sub{tot}$), the peak of the pdf occurs where 
\ba \diff{}{R^*} \phi(R^*, N\sub{tot})= 0. \ea
It follows that the peak of the extreme-value pdf occur at the zero of the function
\ba X(R) = (N-1)f^2 +F \diff{f}{R}.\ea
These zeroes can be used to roughly locate the most likely extreme radius (although when comparing with data in the next section, we will again use the percentiles).

\begin{figure} 
   \centering
   \includegraphics[height=2.5in]{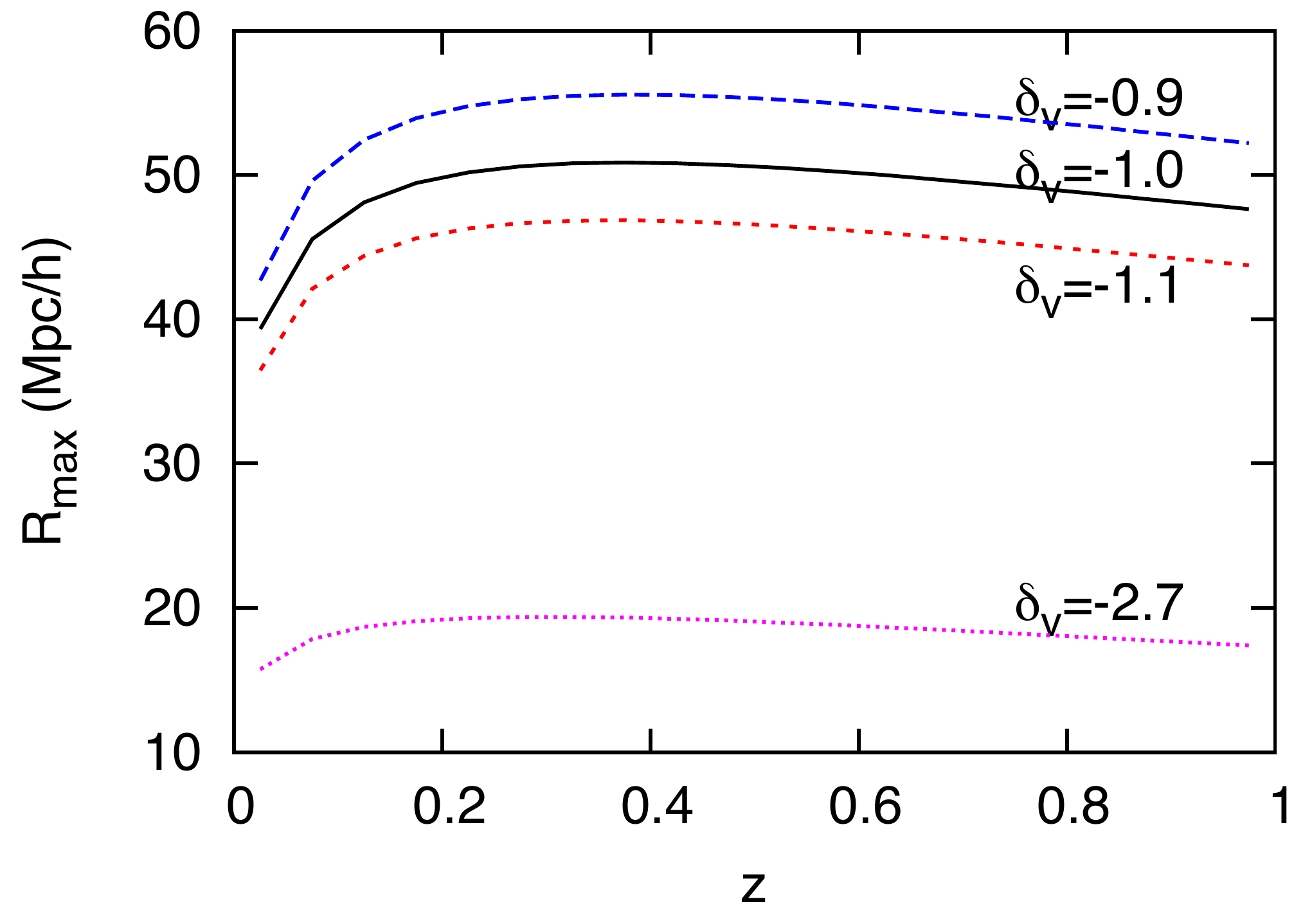}   
   \caption{The distribution of the peak of the pdf, $\phi(R\sub{max})$, where  $R\sub{max}$ is the largest void expected, plotted as a function of redshift. We also show its variation with the void underdensity $\delta_v$. Note that the `shell-crossing' value $\delta_v=-2.7$ predicts an unfeasibly small radius of extreme voids.}
   \label{figextreme}
\end{figure}

Fig. \ref{figextreme} shows $R\sub{max}$, the peak of the extreme-value distribution (\ie{} the root of $X(R)=0$), as a function of redshift. With $\delta_v=-1$, we see that the largest void predicted by the SVdW theory is most likely $\sim50$ $h^{-1}$Mpc in radius, at $z\sim0.2-0.3$, with $R\sub{max}$ very gradually decreasing at higher redshifts. This slow redshift evolution agrees with the simulation of \cite{watson}, in which the maximum void radius appears to be almost independent of redshift.

\subsection{Sensitivity to $\delta_v$}
In Fig \ref{figextreme} , we also plotted the extreme-void predictions for varying $\delta_v$. We observe that decreasing $|\delta_v|$ results in larger extreme-voids across all redshifts, in accordance with the previous section where we deduced that larger voids are shallower.

In addition, we also included the case $\delta_v=-2.7$, the shell-crossing value originally recommended by SVdW. However, such a large magnitude of $|\delta_v|$ predicts the peak of $R\sub{max}$ at only $\sim$ 20$ h^{-1}$Mpc, which is effectively ruled out by recent simulations in which much larger voids have been seen (we will elaborate on this in the next section). Our finding adds weight to the poor fit of $\delta_v=-2.7$ to the void abundance $dn/d\ln R$ as seen in the simulations of \cite{chan}.

\subsection{Sensitivity to dark energy}
The radius of extreme voids is linked to the properties of dark energy.  Here, we consider the effects of varying the dark-energy density parameter, $\Omega_\Lambda$,  and equation of state, $w$, from the fiducial values ($\Omega_\Lambda=0.73, w=-1$).

If $\Omega_\Lambda$ were to increase whilst flatness is maintained, then the matter density parameter must decrease. This gives rise to a reduced $\sigma_8$ and, intuitively, we expect structures to be generally less pronounced; clusters are less massive, and voids are smaller.

If the dark-energy equation of state, $w$, is larger than $-1$, we expect that void radius will not grow as fast at late times, in accordance with Friedmann equation which predicts $R\sim t^{2/(3(w+1))}$. Hence, we expect smaller voids if $w>-1$.

\begin{figure} 
   \centering
   \includegraphics[height=2in]{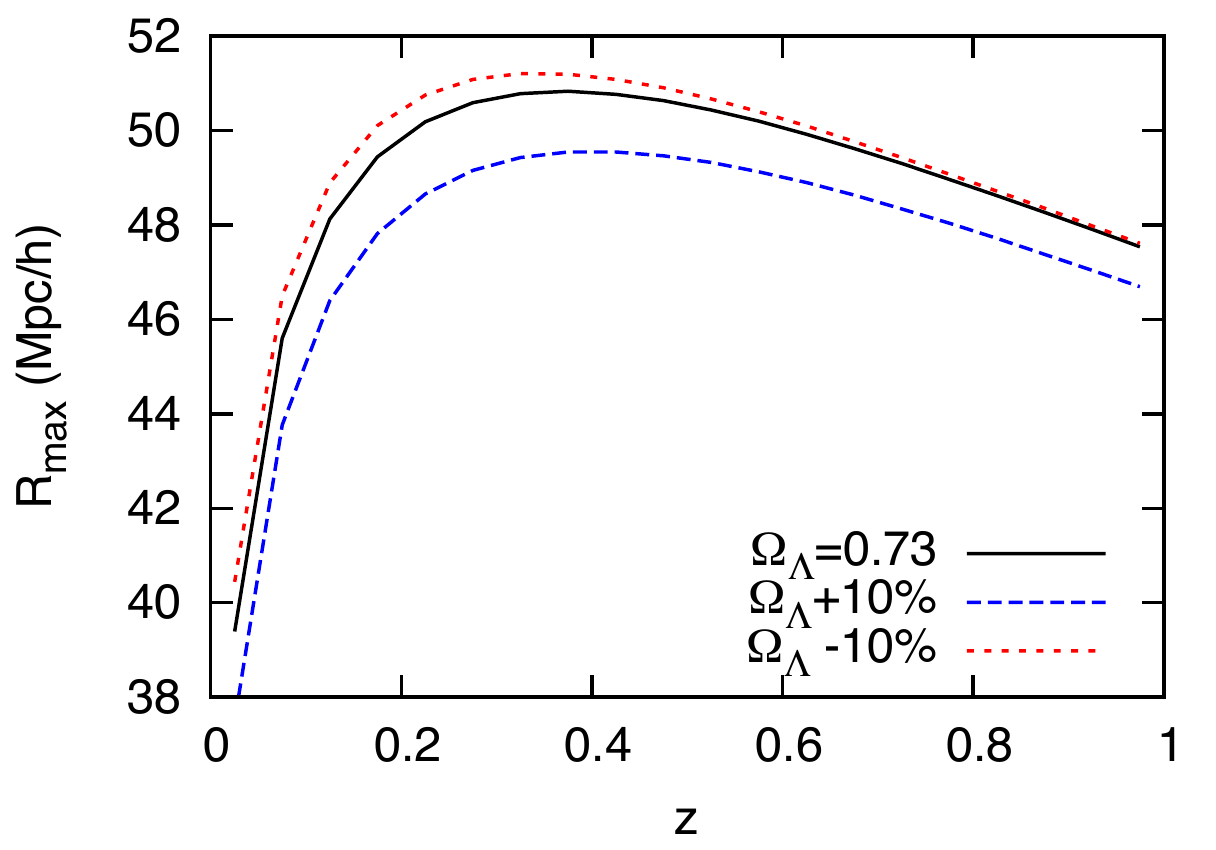} \!\!\!\!\includegraphics[height=2in]{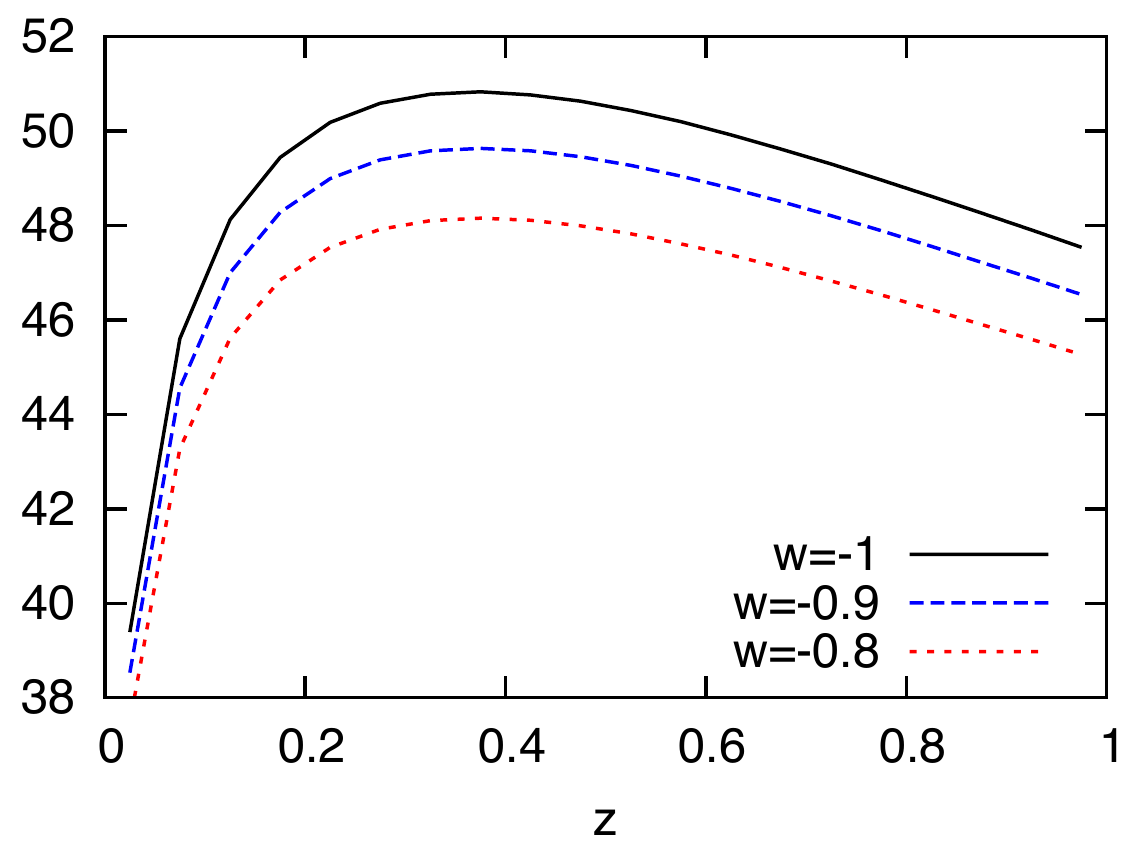}   
   \caption{The sensitivity of $R\sub{max}(z)$ to dark energy. \ii{Left:} The dark-energy density parameter $\Omega_\Lambda$ is varied within 10\% of the fiducial value (whilst flatness is maintained). \ii{Right:} The dark-energy equation-of-state parameter $w$ (assumed redshift independent) is varied from $-0.8$ to $-1.$}
  \label{figde}
\end{figure}

Figure \ref{figde} shows the peak of the extreme-void pdf when $\Omega_\Lambda$ and $w$ are changed from their fiducial values ($\Omega_\Lambda$ is varied whilst flatness is maintained). These behaviours are in accord with our intuitive discussion above. Note that the effect of varying $w$ seen here qualitatively agrees with the results in \cite{pisani}, in which the effects of $w$ on voids are further investigated.  

In both cases, the variations shown are degenerate with small changes in $\delta_v$ in the fiducial model. However, current observational constraints on $\Omega_\Lambda$ and $w$ are so tight that we do not expect these degeneracies to be significant.

\section{Comparison with simulations}\lab{sim}

We now compare our extreme-void model with simulation results. We study the void distributions in two simulations: 1) Chan, Hamaus and Desjacques \cite{chan}, using a 1.5 $h^{-1}$Gpc box, containing $1024^3$ particles, at redshift snapshots $z=0, 0.5$ and 1, 2) Sutter \etal, using a 1 $h^{-1}$Gpc box, containing $1024^3$ particles, at snapshot $z=0$. The maximum void were extracted from these distributions, giving 4 data point in the $z-R\sub{max}$ plane.

On top of these data points, we superimposed the theoretical extreme-value distribution, calculated as in \S\ref{distex}, except that the pdf and the total number of voids are calculated without redshift binning:
\ba
f(R,z)&= {1\over N\sub{tot}} \diff{n}{ R},\\
N\sub{tot}(z)&= \int_{R\sub{min}}^\infty \diff{n}{ R}\,\D R.
\ea

\begin{figure} 
   \centering
   \includegraphics[height=2.5in]{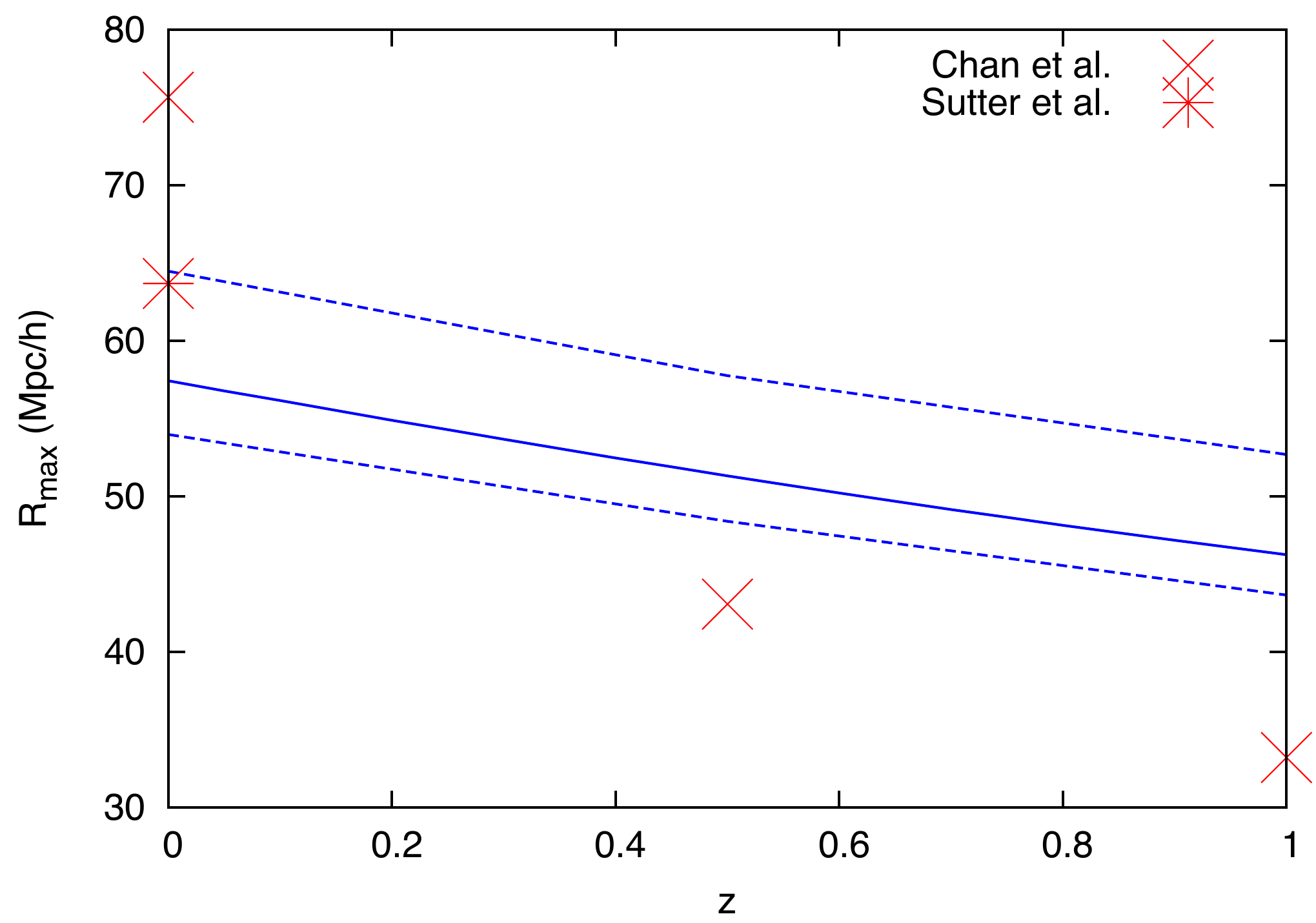}   
   \caption{The radius, $R\sub{max}$, of the largest void computed using the SVdW theory without volume factors. The diagonal stripe comprises the peak of the extreme-value pdf (centre -- solid line), and the 95\% and 5\% percentiles of the distribution (top and bottom dashed lines). Data points (x) are from the simulation of \cite{chan} whilst the point ($*$) is from Sutter \etal \cite{sutter2}.}
   \label{fig_data}
\end{figure}

Fig. \ref{fig_data} shows the result with $\delta_v=-1$ in solid line, with dashed lines representing the 5th and 95th percentiles of the distribution. 

We see that the data from Sutter \etal{} fall within the central 90\% of the distribution. However, those from Chan \etal{} fall significantly outside the contours. Informed by the results in Fig. \ref{figextreme}, it appears that their simulation prefer $\delta_v>-1$ at $z\sim0$, and \ii{decreasing} with redshift. 

This discrepancy from their recommended value of $\delta_v\sim-1$ can be explained by examining their Fig. 2 carefully. Even though the bulk of the void distribution can be described by the SVdW model with $\delta_v\sim-1$, the mismatch at high values of $R$ appears to be significant. 

In conclusion, our results favour a redshift-dependent barrier $\delta_v(z)$ - similar to the modifications of the SVdW model advocated by \cite{paranjape, daloisio}. Although SVdW theory with $\delta_v\approx-1$ appears to fit $dn/d\ln R$ well across the redshifts, we have shown that the data on extreme voids are not consistent (\ie{} outside the central 90\%) with theoretical expectations at any of the redshift snapshots. This illustrates the use of our extreme-value formalism as an additional consistency check for void theories.



The reader may be aware of much larger void radii that have been reported in other simulations and observations. For instance, \cite{watson,nadathur} reported voids as large as 350 $h^{-1}$ Mpc, which, according to the SVdW, should be unobservably rare. However, these voids are not defined using the same criterion as the SVdW definition (in which voids are region with mean density $\rho_v\leq0.2\overline{\rho}$). It is beyond the scope of this work to try to reconcile these two definitions. We simply note here that such enormous voids cannot yet be theoretically described using the SVdW framework without further \ii{ad hoc} adjustments.

\section{Comparison with observation}\lab{sim}

\begin{figure} 
   \centering
   \includegraphics[height=2.5in]{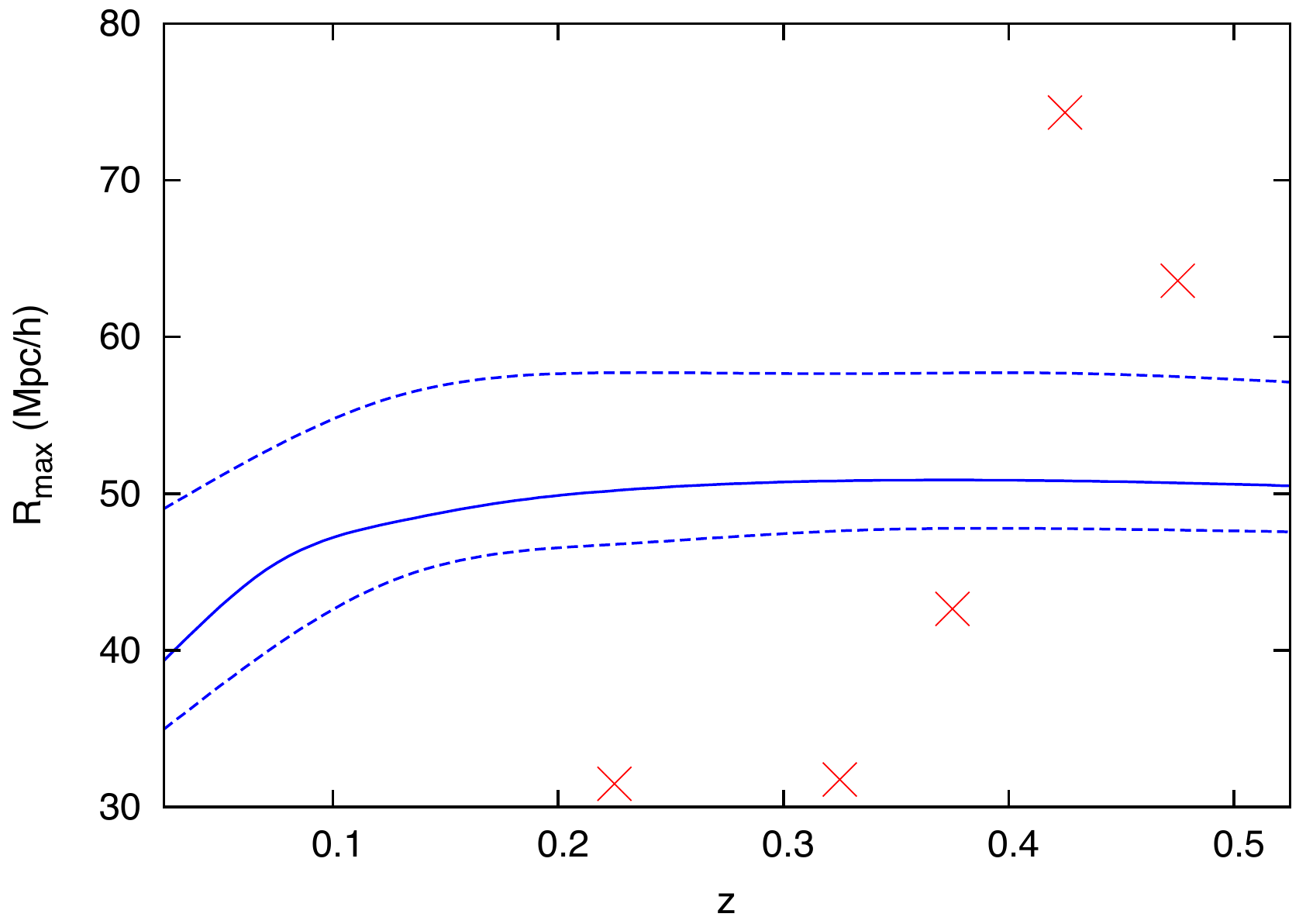}   
   \caption{The radius, $R\sub{max}$, of the largest void expected in each redshift bin with $\Delta z=0.05$ computed using the SVdW theory . The diagonal stripe comprises the peak of the extreme-value pdf (centre -- solid line), and the 95\% and 5\% percentiles of the distribution (top and bottom dashed lines). Data points (x) are from the SDSS10 catalogue \cite{sutter3}.}
   \label{fig_sdss}
\end{figure}

We perform a similar comparison with voids identified from the SDSS10 catalogue as presented in \cite{sutter3}. Voids in the catalogue were binned with $\Delta z=0.05$, and the maximum in each bin is picked out. Changing the bin size has no significant effect on neither the theory nor the observed extreme-void distributions. The result is shown in Fig \ref{fig_sdss}. The theory curve with $\delta_v=-1$ (solid line) is calculated using the formalism in \S\ref{distex}. The 5th and 95th percentiles are shown in dashed lines.

As in the previous section, it appears that no single value of $\delta_v$ is a suitable fit to the observed extreme voids. In contrast with the simulated voids, real voids appear to favour $\delta(z)$ which \ii{increases} with redshift. 

The discrepancy between theory and observation is not surprising: it is unclear how the SVdW predictions, which are relevant only for dark matter voids, can be extrapolated to galaxy voids. Furlanetto and Piran \cite{furlanetto} previously tackled this question by modifying the SVdW theory to fit voids in the 2dF survey. Perhaps a formalism such as theirs is could be used our extreme-value formalism to ameliorate the discrepancy seen in Fig. \ref{fig_sdss}.

\section{Conclusion and discussion}

In summary, we have given a simple theoretical framework for the prediction of the size of the largest voids expected within a given redshift bin. Our model is based on the exact extreme-value calculations previously applied to galaxy clusters. We implement our formalism using the void-abundance model of Sheth and Van de Weygaert, with $\delta_v$ taken as a free parameter. Our results show that the simplest SVdW with $\delta_v=$ a constant is inconsistent with the extreme sizes of voids seen in the simulations and observations. For instance, $\delta_v$ which decreases with redshift would produce a better fit to simulations.

The central message in this work is that any theory which predicts void abundance must also give a sensible prediction for the distribution of \ii{extreme} voids. Our work provides a framework to perform this check.

There are several scopes for improvement of our extreme-value model. 

\bit
\item \ii{Void correlation.} Of course, voids are not completely uncorrelated, as previously shown in the theoretical work of  \cite{white, paranjape,achitouv, chan}. Preliminary works suggest that introducing a void correlation will reduce the SVdW abundance for spherical voids, although the effects on the largest voids are likely to be small and predictable by linear theory \cite{colberg,shethlemson}. Our work on including void correlation into the extreme-void statistics is in progress.

\item \ii{Alternative void abundances.} There is increasing evidence that the SVdW model cannot accurately explain simulation data without \ii{ad hoc} adjustments, simply because real voids are unlikely to be spherical, nor are they all likely to form out of the shell-crossing condition. Apart from simple modification of the SVdW model, alternative void abundances include those directly derived from the Press-Schechter cluster abundance \cite{me, kamionkowski}, or one derived using de-Sitter geometry \cite{gibbons}. The extreme-value formalism presented here remains unchanged and can be readily applied using these alternative models.

A crucial shortfall of the SVdW model is the fact that it assumes that voids are isolated spheres, and as voids evolve from the initial Lagrangian space to the late-time Eulerian space, their volume is not conserved (instead, their comoving number density is conserved -- and only their sizes change). Jennings, Li and Hu \cite{jennings} showed that this assumption can yield volume fraction in voids exceeding unity. They suggested a modified volume-conserving SVdW model -- so-called `$V\, dn$' model, and showed that this formalism produces a good fit to their simulation, but only for spherical voids with $R\leq20h^{-1}$ Mpc (using $\delta_v$=-2.7).

\begin{figure} 
   \centering
   \includegraphics[height=2in]{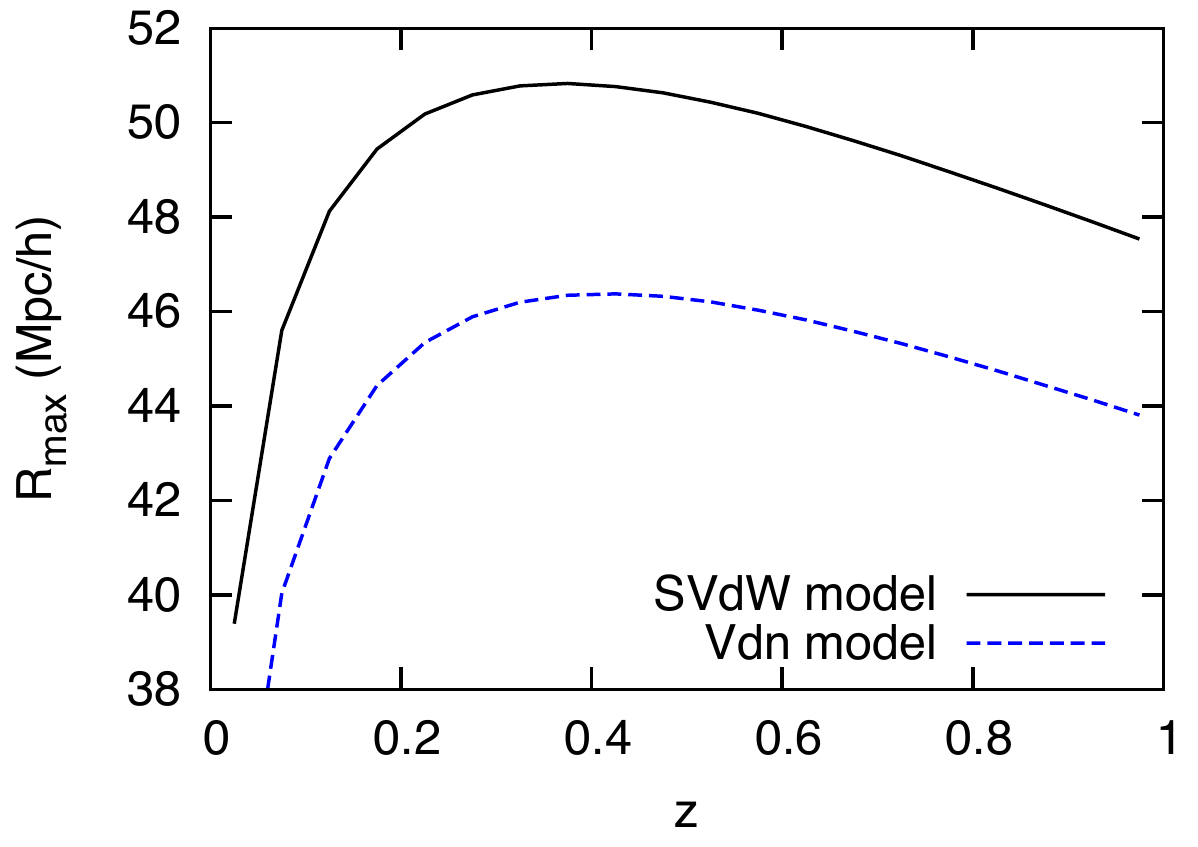} 
   \caption{Comparison between the SVdW model and the volume-conserving `$V$d$n$' model of \cite{jennings}.}
   \label{figvdn}
\end{figure}

It is curious to see if the $V\,dn$ model performs any better than the SVdW model when incorporated into our extreme-value formalism. Figure \ref{figvdn} shows the comparison (with $\delta_v$=-1 in both cases). Unfortunately the $V\,dn$ suppresses the radius of extreme-voids, and exascerbates the underprediction of extreme void radii.

\item \ii{Alternative void definitions.} Voids are defined in the SVdW theory as spherical regions in which the density satisfies $\rho_v\leq0.2\overline{\rho}$ (where $\overline{\rho}$ is the mean matter density). We have found that decreasing the ratio $\rho_v/\overline{\rho}$ to below 0.2 increases the extreme sizes of voids.  However, such an adjustment is unsatisfactory as there are consistency relations between this ratio and $\delta_v$ (see the Appendix of \eg \cite{jennings}).  Even with the consistency relation in place, we were unable to reproduce a significant number of extreme voids with radius beyond $\sim100$ Mpc unless model parameters describe extremely shallow voids. We anticipate that this problem requires a fresh approach to the modelling of void abundance, which must include the shape and profile of voids. 
\eit

Perhaps the greatest challenge for void theory is the prediction for the  abundances of real galaxy voids seen, for instance, in the SDSS data \cite{sdssvoids,sdss2}. A solution to this problem will have a significant implication, in particular, on explaining large-scale observables associated with supervoids. Given the highly nonlinear and complex nature of real voids, perhaps it is more productive to strive towards a semi-analytic modelling of voids led by accurate simulations, similar to what has been achieved with the universal-mass-function approach for galaxy clusters.

\acknowledgments{}
I sincerely thank Chan Kwan Chuen for providing the simulation data, Mark Neyrinck for useful discussions, and Joe Silk for his hospitality at the Institut Astrophysique de Paris, where part of this work was completed. I especially thank the Referee for detailed suggestions which led to significant improvements of the first draft.

\bibliographystyle{jhep}
\bibliography{voids}

\end{document}